\documentclass[prd,twocolumn,showpacs,superscriptaddress,nofootinbib,floatfix,showkeys,10pt]{revtex4-1}
\usepackage{graphicx}
\usepackage{amsmath}
\usepackage{bm}
\usepackage{yhmath}
\usepackage{mathtools}
\usepackage{wasysym}
\usepackage[colorlinks,citecolor=violet,urlcolor=violet,linkcolor=blue]{hyperref}
\usepackage{subfigure}
\usepackage{color}
\usepackage{cases}
\usepackage{subfigure}
\usepackage{times}
\usepackage{dcolumn,booktabs,bm}
\usepackage{slashed}
\usepackage{amsfonts,amssymb,stmaryrd,latexsym,amsmath}
\usepackage{textcomp}
\usepackage{multirow}
\usepackage{cancel}
\usepackage{array}
\usepackage{orcidlink}
\usepackage{xcolor,colortbl}

\def\SU3{{\text{SU(3)}_{\rm F}}}

\def \pcs4338{{P_{\psi s}^\Lambda(4338)^0}}

\allowdisplaybreaks

\begin{document}
	
	\title{\textcolor{violet}{The allowed baryon to baryon-meson strong transitions}}
	
	\author{A. R. Olamaei\,\orcidlink{0000-0003-3529-3002}}\email{olamaei@jahromu.ac.ir}
	\affiliation{Department of Physics, Jahrom University, Jahrom, P.~ O.~ Box 74137-66171, Iran}
	\affiliation{ School of Physics, Institute for Research in Fundamental Sciences (IPM)  P. O. Box 19395-5531, Tehran, Iran}
	
	\author{S.~Rostami\,\orcidlink{0000-0001-7082-6279}}\email{asalrostami.phy@gmail.com}
	\affiliation{Department of Physics, University of Tehran, North Karegar Avenue, Tehran 14395-547, Iran}
	
	\author{K. Azizi\,\orcidlink{0000-0003-3741-2167}}\email{kazem.azizi@ut.ac.ir} \thanks{Corresponding Author}
	\affiliation{Department of Physics, University of Tehran, North Karegar Avenue, Tehran 14395-547, Iran}
	\affiliation{Department of Physics, Dogus University, Dudullu-\"Umraniye, 34775 Istanbul, T\"urkiye}
	
	
	\begin{abstract}
We investigate the kinematically allowed baryon to baryon-meson strong transitions in all the light and heavy sectors.  We consider only  the ground state on-shell particles in the baryonic and mesonic channels.  In the case of mesons,   only the well-established pseudoscslar and vector nonets are involved.   For the baryons,  we consider the ground state spin 1/2 and  3/2 baryons. Using all the restrictions and conservation laws, a \texttt{mathematica}/ \texttt{python}  code selects the allowed strong channels among many possible transitions.  Using the strong coupling constants defining these transitions, we estimate the strong width and beaching fraction at each channel,  that may help ongoing experimental and theoretical   investigations. 
		
	\end{abstract}
	
	\maketitle
	
	\thispagestyle{empty}
	
\textit{\textbf{\textcolor{violet}{Introduction:}}~}
The recent advancements in experimental techniques \cite{Yuan:2021eqz,Rossi:2023xqa} for identifying hadrons containing one, two, or three heavy quarks, coupled with the anticipation of future discoveries, provide us with unprecedented opportunities to investigate their decay processes in a greater detail.
Among these, strong decays hold a special significance. 
Strong decays provide information about the internal structure of hadrons 
(such as protons and neutrons) and how quarks and gluons interact. 
This information can help us develop stronger theories regarding hadronic dynamics. 
Strong decays can be served as laboratories  to test the  existing theories like QCD (Quantum Chromodynamics). 
Analyzing the experimental outcomes of these decay processes allows us to critically assess the validity and limitations of current theoretical models.

Strong decays may lead to the production of new or unstable particles \cite{Barnes:2003bn,Akita:2024nam}. 
Analysing these processes can aid in identifying new particles and examining their characteristics. 
Strong decays can also provide insights into the symmetries present in 
nature and how they break, offering a deeper understanding of fundamental physics. 
Some strong decays can  provide information about the initial conditions of the universe 
after the Big Bang and how hadrons formed during that time.
Strong decays are also  important for the exploration of the hadronization process and the formation of hadronic matter after the Big Bang \cite{Alford:1998mk}. The suppression of heavy quarkonium,  such as $\Upsilon$ and $J/\psi$, in heavy-ion collisions has been associated with the existence of Quark-Gluon Plasma (QGP) \cite{Matsui:1986dk}. Similarly, the production and strong decays of baryons in such environments provide constraints on hadronization models, allowing us to infer the conditions of the early universe \cite{Kohri:2001jx}. Further studies may help our understanding of hadron formation and the role of strong interactions in cosmic evolution.
Overall, the study of strong decays not only enhances our understanding of the structure of matter but also serves as a crucial tool for discovering and validating new theories in fundamental physics. 

In this domain, coupling constants play a fundamental role.
They have the potential to investigate the interactions between 
hadronic configurations, allowing us to better determine the nature 
of strong interactions and their behavior. Up to now, various articles 
have examined and calculated these couplings using different methods \cite{Tawfiq:1998nk,Ivanov:1999bk,Azizi:2008ui,Aliev:2009ei,Huang:2009is,Aliev:2010yx,Aliev:2010ev,Aliev:2010dw,Aliev:2010su,Azizi:2010sy,Detmold:2011bp,Aliev:2011ufa,Azizi:2014bua,Azizi:2015bxa,Azizi:2015jya,Azizi:2015tya,Agaev:2015faa,Aliev:2016qdo,Aliev:2017tej,Yu:2018hnv,Alrebdi:2020rev,Rostami:2020euc,Aliev:2020aon,Aliev:2020lly,Azizi:2020zin,Olamaei:2021hjd,Aliev:2021hqq,Lu:2023pcg}. 
Accurate calculations of these strong coupling constants are essential for determining the decay rates and 
branching ratios to provide useful information for the experiments.
	The understanding of strong interactions among different hadronic configurations  was significantly 
	advanced by recent theoretical and experimental developments such as lattice QCD~\cite{Bahtiyar:2020uuj,Zhang:2021oja},
	QCD sum rules~\cite{Agaev:2017ywp,Aliev:2018vye, Aliev:2018lcs,Yang:2022oog,Agaev:2017lip},
	heavy quark effective theory (HQET)~\cite{Vishwakarma:2022vzy},
	 the Regge phenomenology~\cite{Jia:2019bkr,Oudichhya:2023awb}, and
	the quark models ~\cite{Karliner:2017kfm,Wang:2017kfr,Yang:2017qan,Karliner:2018bms,Shi:2019tji,Karliner:2020fqe,Wang:2020gkn,Xiao:2020gjo,Chen:2021eyk,Garcia-Tecocoatzi:2022zrf,Ma:2022vqf,Wang:2022dmw,Karliner:2023okv,Ortiz-Pacheco:2023bns}.
	These endeavors are further enriched by the contributions of experimental collaborations such as LHCb and Belle, which persistently challenge and refine theoretical models. Notable achievements include the LHCb's observation of five distinct $\Omega_c$ states in $\Xi_c^+ K^-$ decays~\cite{LHCb:2017uwr, Belle:2017ext} and the subsequent identification of $\Lambda_c$, $\Xi_c$, and bottom baryons ($\Omega_b$, $\Xi_b$, $\Lambda_b$) over the period from 2018 to 2022~\cite{LHCb:2018haf, LHCb:2020tqd, LHCb:2021ssn}. Additionally, the observation of the decay $\Xi_b^- \to \Lambda_b^0 \pi^-$ \cite{LHCb:2023tma} stands as a significant milestone, further advancing our understanding of baryonic processes and their implications within the broader framework of hadronic physics.
	The precise and reliable extraction of coupling constants remains a cornerstone for addressing persisting ambiguities in hadron spectroscopy. Such advancements are crucial for deepening our comprehension of QCD within its non-perturbative regime, thereby shedding light on the underlying mechanisms that govern hadronic structures and interactions.

Among various methods,  the QCD sum rule approach \cite{Shifman:1978bx} has a special place since it is based on the QCD Lagrangian and enables us to calculate observables like strong couplings analytically \cite{Colangelo:2000dp}. Then one can use these couplings to calculate transition amplitudes and decay rates for various strong decays.

In the present work, by implementing the strong couplings calculated via the QCD sum rules method, we investigate all possible strong decays of baryons into mesons and baryons, and calculate their corresponding decay rates. Although it appears that numerous strong transitions are possible, conservation laws severely constrain their numbers. This analysis can be used for furthur experimental investigations which can be used to determine the strong properties of harons more accurately. 

In the rest of this letter, first we briefly describe the method to calculate the strong couplings and then calculate the matrix elements and decay rates for all kinds of strong decays among baryons and mesons. Then by implementing the conservation laws we explore all allowed decays and calculate their corresponding decay rates and branching fractions.

\textit{\textbf{\textcolor{violet}{Physical Observables:}}~} 
QCD sum rule is one of the most successful methods to investigate the strong, weak and electromagnetic transitions of hadrons as well as their properties like mass and widths analytically. It is based on QCD Lagrangian 
\begin{eqnarray}\label{eq:QCDLag}
	{\cal L}_{\text{QCD}} = \bar{\psi}_i (i \slashed{D}_{ij} - m_i \delta_{ij}) \psi_j - \frac{1}{4} G^{a}_{\mu\nu}G^{\mu\nu}_{a}~,
\end{eqnarray}
and can be used in both perturbative and non-perturbative realms.  In above equation  $ \psi_i $ is the quark field, a dynamical function of spacetime,  $ G^{a}_{\mu\nu}  $ is gluon field strength tensor  and  $ (D_\mu)_{ij} $ is the gauge covariant derivative. The starting point is to write the corresponding correlation function (CF)
\begin{eqnarray} \label{eq:CorrF1}
	\Pi (p, q) = i \int d^4 x e^{ip.x} \langle {\cal M}(q)|\mathcal{T}\{\eta_2(x)\bar{\eta}_1(0)\}|0\rangle~,
\end{eqnarray}
where ${\cal T}$ is the time ordered operator, $\eta_1$ and $\eta_2$ are the interpolating currents representing the initial and final baryons, ${\cal B}_1(p+q)$ and ${\cal B}_2(p)$, respectively and ${\cal M}(q)$ stands for the final meson. The interpolating currents must contain the corresponding baryon quantum numbers as well as its Lorentz structure.  In the light-cone version of the sum rule method, the distribution amplitudes (DAs) of the on-shell meson $ {\cal M}  $ is used.

To calculate the strong couplings we insert the complete set of hadronic states into the CF which gives us the real part of the CF in the time-like region of the light-cone as
\begin{eqnarray}\label{eq:QCDofCF}
	\Pi(p, q) &=& \frac{1}{(p^2 - m _{\mathcal B _2}^2) ((p + q)^2 - m _{\mathcal B _1}^2)} \nonumber \\ 
	&\times& \langle{0}|{\eta_2}|{\mathcal B _2(p)}\rangle \langle{\mathcal M(q) \mathcal B_2 (p)}|{\mathcal B_1 (p + q)}\rangle\nonumber \\ 
	&\times& \langle \mathcal B_1(p + q) |\bar{\eta_1}|0\rangle \nonumber \\
	& + & \mbox{~higher states and continuum}~, 
\end{eqnarray}
including  the matrix elements  $\langle{\mathcal M(q) \mathcal B_2 (p)}|{\mathcal B_1 (p + q)}\rangle  $,  which are parametrized in terms of strong coupling form factors,  $ g_i(q^2) $. The strong coupling form factors at $ q^2=m^2_{\mathcal M} $ is called the strong coupling constants,  $ g_i $. The remaining matrix elements are defined in terms of the residues of the participating baryons.  The resultant expression  is the physical side of the CF in terms of the strong form factors, residues and different Lorentz structures.   On the other hand, by inserting the corresponding interpolating currents into the CF and using Wick theorem to eliminate the time ordering operator,  as well as performing operator product expansion to separate the perturbative and non-perturbative parts of the CF,  one can write the CF in terms of light and heavy quark propagators and DAs of the corresponding final meson. The final result, which is found after Fourier and Borel transformations as well as continuum subtractions,  is the imaginary part of the CF which is the QCD or theoretical side of it in the space-like region of the light cone. To connect the physical side of the CF which is the real part of it to the QCD side which is the the imaginary part, we use the dispersion integral
\begin{eqnarray}\label{eq:dispint}
	\Pi(p^2, q^2)= \int \frac{1}{\pi} \frac{\text{Im}\big\{\Pi(s, t,q^2)\big\}}{(s - p^2)(t - q^2)} ds dt~,
\end{eqnarray}
that connect the real and imaginary parts of any analytical complex function for the coefficient of each Lorentz structure for the ground state particles.
The integral is divergent on the upper limits and therefore to increase the radius of convergence, as said above,  we use the Borel transformation on both sides as well as continuum subtraction supplied by the quark-hadron duality assumption,  to suppress the contributions of the excited and continuum states and enhance the ground state contribution. The final result is the strong coupling constants in terms of QCD  fundamental parameters and some auxiliary quantities.  The auxiliary parameters such as Borel parameters and continuum threshold are fixed based on the standard prescriptions of the method. These couplings are used as main inputs to calculate the matrix elements and decay rates of the corresponding strong decay of baryon ${\cal B}_1$ to baryon ${\cal B}_2$ and meson ${\cal M}$.




\textit{\textbf{\textcolor{violet}{Decay Rates:}}~}
If the amplitude of ${\cal B}_1 \rightarrow {\cal B}_2 {\cal M}$ is denoted by $M$, the corresponding decay rate, $\Gamma$, would be:
\begin{eqnarray}\label{eq:decayrate}
	\Gamma = \frac{p_f}{8 \pi m_{{\cal B}_1}^2 (2s+1)} |M|^2~,
\end{eqnarray}
where $s$ is the spin of the initial baryon ${\cal B}_1$ and $p_f$ is the momentum of any final hadrons in the rest frame of ${\cal B}_1$ which is
\begin{eqnarray}\label{eq:pf}
	p_f = \frac{\sqrt{\lambda(m_{{\cal B}_1}, m_{{\cal B}_2}, m_{{\cal M}})}}{2 m_{{\cal B}_1}}~,
\end{eqnarray}
where 
\begin{eqnarray}
	\lambda(m_1, m_2, m_3) &=& m_{1}^4 +m_{2}^4 +m_{3}^4 -2 m_{1}^2 m_{2}^2 \nonumber \\
	&&-2 m_{1}^2 m_{3}^2 -2 m_{2}^2 m_{3}^2~.
\end{eqnarray}
The number of couplings depends on the spin of hadrons involved in the decay.
 Considering the constraints dictated by conservation laws in decay processes, two key scenarios can be identified. In both, the final baryon and meson possess spins of $\frac{1}{2}$ and $0$, respectively, while the initial baryon exhibits a spin of either $\frac{1}{2}$ or $\frac{3}{2}$.

For the case $s_{{\cal B}_1} = 1/2$, $s_{{\cal B}_2} = 1/2$ and $s_{\cal M} = 0^-$, the decay rate is
\begin{eqnarray}\label{decayrate1212p}
	\Gamma = \frac{g^2~ \Big[(m_{{\cal B}_1} - m_{{\cal B}_2})^2-m_{{\cal M}}^2\Big] }{16 \pi  m_{{\cal B}_1}^3} \sqrt{\lambda(m_{{\cal B}_1}, m_{{\cal B}_2}, m_{{\cal M}})}~.
\end{eqnarray}
where the sum over the initial and final baryon spin is performed using
\begin{eqnarray}
	\sum _s u(p,s) \bar u(p,s) = \slashed p + m~,
\end{eqnarray}
and $u(p,s)$ is the spinor with momentum $p$ and spin $s$. 
For the case $s_{{\cal B}_1} = 3/2$, $s_{{\cal B}_2} = 1/2$ and $s_{\cal M} = 0^-$ the decay rate is
\begin{eqnarray}
	\Gamma &=& \frac{g^2 \sqrt{\lambda(m_{{\cal B}_1}, m_{{\cal B}_2}, m_{{\cal M}})}}{192 \pi  m_{{\cal B}_1}^5} \\
	&&\times  \Big[ (m1 - m2)^2 - m3^2 \Big]^2 \Big[ (m1 + m2)^2 -m3^2 \Big], \nonumber
\end{eqnarray}
where for the baryons with spin 3/2, the spin sum is
\begin{eqnarray}\label{eq:spin3/2sum}
	\sum _s u _\mu (p,s) \bar u _\nu (p,s) &=& -(\slashed p + m) \Big[
	g _{\mu\nu} - \frac 13 \gamma _\mu \gamma _\nu \\
	&&- \frac 23 \frac{p_\mu p_\nu}{m^2} +\frac 13 \frac{p_\mu \gamma_\nu - p_\nu \gamma _\mu} m
	\Big]~,\nonumber
\end{eqnarray}
and $u _\mu (p,s)$ is the Rarita–Schwinger spinor.

 In both cases, there are only one strong coupling $g$ involved due to the simple Lorentzian structure of the hadrons. The corresponding decay rate using, for instance \ref{eq:decayrate},  is
\begin{eqnarray}\label{decayrate1212p}
	\Gamma = \frac{g^2~ \Big[(m_{{\cal B}_1} - m_{{\cal B}_2})^2-m_{{\cal M}}^2\Big] }{16 \pi  m_{{\cal B}_1}^3} \sqrt{\lambda(m_{{\cal B}_1}, m_{{\cal B}_2}, m_{{\cal M}})}~.
\end{eqnarray}
Having $g$ from QCD sum rule calculations, one can evaluate the decay rate $\Gamma$ for the corresponding transition.



\textit{\textbf{\textcolor{violet}{Analysis:}}~}
In this part of the analysis, we investigate the probability 
of strong decay of a baryon to a baryon and meson. 
Initially, we consider 46 different mesons (pseudoscalar and vector)  and 50 baryon states for each spin 3/2 and 1/2, 
leading to a total of $50 \times 46 \times 50 = 115000 $ potential decay scenarios. But the conservation laws severely restrict the number of possible decays.

The first conservation law to take into account is the kinematical energy-momentum conservation. 
In the rest frame of initial baryon,  this means that the combined mass of the final baryon and meson must be 
less than that of the initial baryon. Applying this constrain narrows down 
the possibilities to 23281 valid states.
Next, we apply the total angular momentum conversations.  
To illustrate this, we begin with initial spin $\frac{3}{2}$ particles. They have no associated orbital angular momentum,  implying $L = 0$. Since only the total angular momentum $J$ is conserved (not the individual components $L$ and spin $S$), the final state must have the same total $J$ as the initial state: $J = \frac{3}{2}$.  In the final state, the spin $\frac{1}{2}$ particles and the mesons are distinct particles, and they can possess a non-zero relative orbital angular momentum $L$. The combined angular momentum arises from summing $L$ with the  spin of $\frac{1}{2}$.  Hence,  for the total angular momentum to be $\frac{3}{2}$, $L$ must be either $ 1 $ or $ 2 $ as both $1 + \frac{1}{2}$ and $2 - \frac{1}{2}$ can give $\frac{3}{2}$.  Next, we consider the parity. The initial spin $\frac{3}{2}$ particles have  positive parity.  For the final state (baryon+meson)  system, the total parity is determined by multiplying their intrinsic parities ($+1$ for the final baryon and $-1$ for the final meson) and the spatial factor $(-1)^L$ associated with their orbital motion. Since parity is conserved in strong interactions, the total parity must remain positive. This condition can only be satisfied if $L$ is odd. Consequently, $L = 2$ is not allowed, and the only acceptable value is $L = 1$.

Considering the above remarks,  for kinematically allowed decays, two distinct scenarios arise based on the spins of the initial and final baryons in the ground states:
\begin{itemize}
	\item \textbf{$\Delta S=0$ transitions}: The spin of the initial and final states is equal.
	\item \textbf{$\Delta S=1$ transitions}: The spin of the initial state exceeds that of the final state by one unit.
\end{itemize}

The final condition pertains to the quark structure of the resulting hadrons based on the fact that strong decays do not change the flavor of the quarks involved which means if the quark content of the initial and final baryons is the same, 
the meson should have the structure $q\bar{q}$ (the same quark-antiquark). However, 
if the baryon structures are different, the baryons must have two identical quarks, 
and the quark structure of the meson inherits different quarks: a quark from the 
initial baryon and an antiquark with the same flavor from the final baryon.

Considering all these constraints,  for the first case, $\Delta S=0$,  only  five possible outcomes could occur:
\begin{eqnarray}
	\Sigma _c{}^{\text{++}}\rightarrow & \Lambda _c{}^+ & \pi ^+, \\ \nonumber
	\Sigma _c{}^+ \rightarrow & \Lambda _c{}^+ & \pi ^0, \\ \nonumber
	\Sigma _c{}^0 \rightarrow & \Lambda _c{}^+ & \pi ^-, \\ \nonumber
	\Sigma _b{}^+ \rightarrow & \Lambda _b{}^0 & \pi ^+, \\ \nonumber
	\Sigma _b{}^-\rightarrow  & \Lambda _b{}^0 & \pi ^-.
\end{eqnarray}
From the \textit{ iso-spin symmetry} that we use  it is obvious that the decay constants for the first three and last two should be the same. The final decay rates are obviously different due to different kinematical properties of the baryons involved.
The results for the corresponding strong decay constants and decay rates are presented in Table \ref{tab:channels}. 
From Table \ref{tab:channels}, it is evident that the heavier the baryon involved, the greater their couplings and strong decay rates. These results indicate shorter lifetimes of the heavier states,  and are  consistent with the physical expectations.
The calculated branching ratios, within the uncertainties,  are consistent with the Particle Data Group (PDG)'s estimates \cite{ParticleDataGroup:2024cfk}, indicating that the corresponding strong decays are  the dominant channels.

For the $\Delta S=1$ case, there are 33 allowed decays, which are presented in Table \ref{tab:decay_channels}. Some of the corresponding strong couplings are computed using the QCD sum rules in Refs.~\cite{Aliev:2010ev,Aliev:2010su}, while the others can be derived from these results by applying $SU(2)_f$ flavor symmetry and utilizing the relations between the corresponding CF's.
Some comments on this issue are in order: 
The sum rules for the strong decay of ${\cal B}_1 \rightarrow {\cal B}_2 {\cal M}$ where ${\cal M}$ is a pseudoscalar meson is
\begin{eqnarray}\label{eq:gsumrule}
	g_{{\cal B}_1 {\cal B}_2 {\cal M}} = \dfrac{e^{m_{{\cal B}_1}^2 / M_1^2 + m_{{\cal B}_2}^2 / M_2^2 + m_{{\cal M}}^2 / (M_1^2 + M_2^2)}}{m_{{\cal B}_2} \lambda_{{\cal B}_1} \lambda_{{\cal B}_2}} \Pi_{{\cal B}_1 {\cal B}_2 {\cal M}},\nonumber\\
\end{eqnarray}
where $\Pi_{{\cal B}_1 {\cal B}_2 {\cal M}}$, $\lambda_{i}$ and $M_i$ are the invariant function of the decay, baryon residue and Borel parameter respectively.
Having determined the strong coupling for a specific decay, one can compute the strong couplings of several analogous decays by utilizing the relations among the corresponding CF's and applying symmetries such as $SU(2)_f$.  
Now,  we present the detailed calculations for several representative decays.
From Eq. \ref{eq:gsumrule} the ratio of two strong couplings correspond to ${\cal B}_1 \rightarrow {\cal B}_2 {\cal M}$ and ${\cal B}'_1 \rightarrow {\cal B}'_2 {\cal M}'$ decays is
\begin{eqnarray}\label{eq:g-ratio}
	\dfrac{g_{{\cal B}_1 {\cal B}_2 {\cal M}}}{g_{{\cal B}'_1 {\cal B}'_2 {\cal M}'}} &=& \dfrac{m_{{\cal B}'_2} \lambda_{{\cal B}'_1} \lambda_{{\cal B}'_2}}{m_{{\cal B}_2} \lambda_{{\cal B}_1} \lambda_{{\cal B}_2}} \cdot \dfrac{\Pi_{{\cal B}_1 {\cal B}_2 {\cal M}}}{\Pi_{{\cal B}'_1 {\cal B}'_2 {\cal M}'}} \\
	&& \times \dfrac{e^{m_{{\cal B}_1}^2 / M_1^2 + m_{{\cal B}_2}^2 / M_2^2 + m_{{\cal M}}^2 / (M_1^2 + M_2^2)}}{e^{m_{{\cal B}'_1}^2 / M_1^{'2} + m_{{\cal B}'_2}^2 / M_2^{'2} + m_{{\cal M}'}^2 / (M_1^{'2} + M_2^{'2})}}. \nonumber
\end{eqnarray}
For baryons belonging to the same category, such as $\Delta$ baryons $(\Delta^{++}, \Delta^{+}, \Delta^{0}, \Delta^{-})$ or $\Sigma^*$ baryons $(\Sigma^{*+}, \Sigma^{*0}, \Sigma^{*-})$, the mass splitting is less than $1\%$. Consequently, their corresponding Borel parameters and residues are identical due to the $SU(2)_f$ flavor symmetry.
Therefore, the ratio of the strong couplings, where the initial baryons as well as the final baryons and mesons belong to the same category, is identical to the ratio of the corresponding invariant functions:
\begin{eqnarray}\label{eq:g-ratio2}
	\dfrac{g_{{\cal B}_1 {\cal B}_2 {\cal M}}}{g_{{\cal B}'_1 {\cal B}'_2 {\cal M}'}} = \dfrac{\Pi_{{\cal B}_1 {\cal B}_2 {\cal M}}}{\Pi_{{\cal B}'_1 {\cal B}'_2 {\cal M}'}}~,
\end{eqnarray}
and thus, knowing one strong coupling, one can find others by just evaluating the ratio of corresponding invariant functions.

The relation between various kinds of invariant functions are calculated, for instance,  in Refs. \cite{Aliev:2010yx,Aliev:2010ev, Aliev:2010dw,Aliev:2010su}.Here, we present the detailed calculations for the strong decay $\Delta \rightarrow N \pi$ as an example. The remaining decays can be treated using the same procedure:

From Table 3 of \cite{Aliev:2010su}, it is known that $g_{\Delta^0 p \pi^-} = 5.5 \pm 1.5$. Using this value, other strong couplings can be determined by calculating the ratio of invariant functions through the relation presented in Ref.~\cite{Aliev:2010su}, as detailed below. For $\Delta^- \rightarrow n \pi^-$ and  $\Delta^0 \rightarrow p \pi^-$ we have
\begin{eqnarray}
	\Pi_{\Delta^- n \pi^-} &=& 2 \sqrt{3} \Pi_1(d, d, d)~,\nonumber\\
	\Pi_{\Delta^0 p \pi^-} &=& 2 \Pi_1(u, u, d)~,
\end{eqnarray}
which form \ref{eq:g-ratio2} follows
\begin{eqnarray}
	g_{\Delta^- n \pi^-} = \sqrt{3}~ \dfrac{\Pi_1(d, d, d)}{\Pi_1(u, u, d)} ~g_{\Delta^0 p \pi^-}~. 
\end{eqnarray}
In the $SU(2)_f$ symmetry limit we have
\begin{eqnarray}
	\Pi_1(d, d, d) = \Pi_1(u, u, d)~,
\end{eqnarray}
where leads to
\begin{eqnarray}
	g_{\Delta^- n \pi^-} = \sqrt{3} g_{\Delta^0 p \pi^-} = 9.5 \pm 2.6~.
\end{eqnarray}
For $ \Delta^0 \rightarrow n \pi^0 $ we have
\begin{eqnarray}
	\Pi_{\Delta^0 n \pi^0 } = \sqrt{2} \Pi_1 (d, d, u) - \dfrac{1}{\sqrt{2}} \Pi_2 (d, d, u).
\end{eqnarray}
Considering the relation
\begin{eqnarray}
	\Pi_2 (u, d, s) = -\Pi_1 (s, u, d) - \Pi_1 (s, d, u)
\end{eqnarray}
and $SU(2)_f$ symmetry one finds
\begin{eqnarray}
	\Pi_{\Delta^0 n \pi^0} = 2 \sqrt{2} \Pi_1 (d, d, u)~,
\end{eqnarray}
which leads to
\begin{eqnarray}
	g_{\Delta^0 n \pi^0} = \sqrt{\dfrac{2}{3}} ~ g_{\Delta^- n \pi^-} = 7.8 \pm 2.1~.
\end{eqnarray}
Similarly one can finds that $g_{\Delta^{++} p \pi^-} = 9.5 \pm 2.6$, $g_{\Delta^{+} p \pi^0} = 7.8 \pm 2.1$ and $g_{\Delta^+ n \pi^+} = 5.5 \pm 1.5~$.
The remaining strong couplings can be calculated using a similar approach. Their corresponding decay rates and branching ratios are presented in Table \ref{tab:decay_channels}.  Note that,  as it is clear from the this table, we consider the electric charges and quark contents of all the initial and final particles explicitly. The existing results in PDG do not contain such details. For instance in the case of $ \Delta \rightarrow  N  \pi $, even the widths/branching ratios for neutron and proton are not separated.  Hence, it is not possible to compare the results for some channels with those of the PDG safely.    As mentioned,  this is mainly due to the fact that, in our approach, we  explicitly consider  all the  isospin components $I_3$  of the initial and final hadrons due to their quark contents.
 For instance, in decays like $ \Delta \rightarrow N \pi$, discussed above, 
	we provide all the isospin components for all charge configurations of the involving initial baryon ($\Delta^{++}, \Delta^{+}, \Delta^{0}$ and $\Delta^{-}$), final baryon ($p^+$ and $n^0$) and final meson ($ \pi^+,  \pi^0,$ and $  \pi^-$), whereas in the PDG such modes are usually reported collectively without charge separation. However, when the contributions from all relevant isospin components are properly summed,  our results demonstrate a meaningful and qualitatively consistent agreement with the PDG values, within the stated uncertainties.   Future experimental analyses with higher resolution in charge-specific channels could further test and validate our detailed predictions.

\begin{table}
	\caption{The strong couplings,  and the corresponding decay widths and branching fractions  for the allowed strong decays of ${\cal B}_1 \rightarrow {\cal B}_2 {\cal M}$ for $\Delta S=0$ .}  \label{tab:channels} 
	\label{tab:example}
	\small
	\centering
	\begin{tabular}{lcccr}
		\toprule\toprule
		\hline
		\textit{Channel}         &  $g$ \cite{Azizi:2008ui,Aliev:2010yx}  & $\Gamma$ $(\text{MeV})$& $Br $\\
		\midrule
		\hline \hline
		$ \Sigma_c^{++}\rightarrow  \Lambda_c^+  \pi ^+ $    &   $6.5 \pm2.4$  & $1.38^{+1.00}_{-0.88}$ & $0.75^{+0.64}_{-0.49}$\\
		\hline
		$  \Sigma_c^+ \rightarrow  \Lambda_c^+ \pi ^0$    &   $6.5  \pm 2.4 $ & $1.35^{+0.99}_{-0.86}$& $0.60^{+0.62}_{-0.42}$  \\
		\hline
		$  \Sigma_c^0 \rightarrow  \Lambda_c^+  \pi ^- $    &  $6.5  \pm 2.4 $ & $1.38^{+1.00}_{-0.88}$   &$0.77^{+0.68}_{-0.51}$\\
		\hline
		$ \Sigma_b^+ \rightarrow  \Lambda_b^0  \pi ^+ $    &   $23.5  \pm 4.9 $  & $4.59^{+1.97}_{-1.81}$& $0.92^{+0.52}_{-0.42}$  \\
		\hline
		$ \Sigma_b^-\rightarrow   \Lambda_b^0   \pi ^-$    &   $23.5   \pm 4.9$  & $4.95^{+2.12}_{-1.96} $& $0.94^{+0.52}_{-0.42}$ \\
		\hline
		\bottomrule
	\end{tabular}
\end{table}

\begin{table*}[htbp]
	\centering
	\small
	\begin{minipage}[t]{0.48\textwidth}
		\centering
		\begin{tabular}{lccc}
			\hline
			\textit{Channel} & $g (\text{GeV})^{-1} $ & $\Gamma$ (MeV) & $Br$ \\
			\midrule
			\hline \hline
			$\Delta^{++} \rightarrow \pi^+ + p$ & $9.5 \pm 2.6 $ & $ 45.5 \pm 24.8 $ & $0.41 \pm 0.22$ \\
			\hline
			$\Delta^+ \rightarrow \pi^+ + n$ &$5.5 \pm 1.5$ & $ 15.0 \pm 8.2 $ & $0.11 \pm 0.06$\\
			\hline
			$\Delta^+ \rightarrow \pi^0 + p$ & $7.8 \pm 2.1$ & $31.5 \pm 16.9$ & $0.24 \pm 0.13$\\
			\hline
			$\Delta^0 \rightarrow \pi^0 + n$ & $7.8 \pm 2.1$ & $ 30.1 \pm 16.7 $ & $0.27 \pm 0.15$ \\
			\hline
			$\Delta^0 \rightarrow \pi^- + p$ &$5.5\pm 1.5$ & $ 15.3 \pm 8.3 $ & $0.14 \pm 0.08$ \\
			\hline
			$\Delta^- \rightarrow \pi^- + n$ &$9.5 \pm2.6 $ & $ 44.8 \pm 24.5 $ & $0.34 \pm 0.19$\\
			\toprule
			\midrule
			\hline\hline
			$\Sigma^{*+} \rightarrow \pi^0 + \Sigma^+$ &$3.4 \pm 0.5$ & $ 1.1 \pm 0,3 $ & $0.03 \pm 0.01$ \\
			\hline
			$\Sigma^{*+} \rightarrow \pi^+ + \Sigma^0$ &$3.4 \pm 0.5$ & $ 0.9 \pm 0.3 $ & $0.02 \pm 0.01$ \\
			\hline
			$\Sigma^{*0} \rightarrow \pi^0 + \Sigma^0$ & $3.4 \pm 0.5 $ & $ 0.8 \pm 0.2 $ & $0.02 \pm 0.01$ \\
			\hline
			$\Sigma^{*0} \rightarrow \pi^+ + \Sigma^-$ &$3.4 \pm 0.5 $ & $ 0.8 \pm 0.2 $ & $0.02 \pm 0.01$ \\
			\hline
			$\Sigma^{*0} \rightarrow \pi^- + \Sigma^+$ &$3.4 \pm 0.5$ & $ 1.0 \pm 0.3 $ & $0.03 \pm 0.01$ \\
			\hline
			$\Sigma^{*-} \rightarrow \pi^0 + \Sigma^-$ &$3.4 \pm 0.5 $ & $1.0 \pm 0.3$ & $0.03 \pm 0.01$ \\
			\hline
			$\Sigma^{*-} \rightarrow \pi^- + \Sigma^0$ &$3.4 \pm 0.5 $ & $ 1.1 \pm 0.3 $ & $0.03 \pm 0.01$\\
			\hline \hline
			$\Xi^{*0} \rightarrow \pi^0 + \Xi^0$ &$3.3 \pm 0.7$ & $ 2.0 \pm 0.9 $ & $0.22 \pm 0.11$ \\
			\hline
			$\Xi^{*0} \rightarrow \pi^+ + \Xi^-$ &$4.7 \pm1.0 $ & $3.3 \pm 1.4$ & $0.36 \pm 0.17$\\
			\hline
			$\Xi^{*-} \rightarrow \pi^- + \Xi^0$ &$4.7 \pm 1.0$ & $ 4.1 \pm 1.7 $ & $0.41 \pm 0.25$ \\
			\hline
			$\Xi^{*-} \rightarrow \pi^0 + \Xi^-$ & $3.3 \pm 0.7$& $ 1.9 \pm 0.8 $ & $0.19 \pm 0.12$ \\
		\end{tabular}
	\end{minipage}
	\hfill
	\begin{minipage}[t]{0.48\textwidth}
		\centering
		\begin{tabular}{lccc}
			\hline
			\textit{Channel} & $g (\text{GeV})^{-1} $ & $\Gamma$ (MeV) & $Br$ \\
			\midrule
			\hline \hline
			$\Sigma^{*+} \rightarrow \pi^+ + \Lambda^0$ &$7.0 \pm 1.5$ & $ 18.9 \pm 8.1 $ & $0.52 \pm 0.23$ \\
			\hline
			$\Sigma^{*0} \rightarrow \pi^0 + \Lambda^0$ & $7.0 \pm 1.5$& $ 19.8 \pm 8.5 $ & $0.55 \pm 0.31$ \\
			\hline
			$\Sigma^{*-} \rightarrow \pi^- + \Lambda^0$ &$7.0 \pm 1.5$ & $ 20.1 \pm 8.6 $ & $0.51 \pm 0.25$\\
			\hline \hline
			$\Sigma_c^{*++} \rightarrow \pi^+ + \Lambda_c^+$ &$7.8 \pm 1.0$ & $16.4 \pm 4.2$ & $ 1.0 \pm 0.3 $ \\
			\hline
			$\Sigma_c^{*+} \rightarrow \pi^0 + \Lambda_c^+$ &$ 7.8\pm 1.0$ & $ 17.0 \pm 4.4 $ & $0.99 \pm 0.49$\\
			\hline
			$\Sigma_c^{*0} \rightarrow \pi^- + \Lambda_c^+$ &$ 6.5\pm 2.4$ & $ 1.6 \pm 1.2 $ & $0.10 \pm 0.08$\\
			\hline \hline
			$\Xi_c^{*+} \rightarrow \pi^0 + \Xi_c^+$ &$3.1 \pm 1.1$ & $ 0.7 \pm 0.5 $ &  $0.33 \pm 0.26$ \\
			\hline
			$\Xi_c^{*+} \rightarrow \pi^+ + \Xi_c^0$ &$4.4 \pm 1.6$ & $ 1.1 \pm 0.8 $ & $0.51 \pm 0.42$\\
			\hline
			$\Xi_c^{*0} \rightarrow \pi^0 + \Xi_c^0$ &$3.1 \pm 1.1$ & $ 0.7 \pm 0.5 $ & $0.30 \pm 0.24$ \\
			\hline
			$\Xi_c^{*0} \rightarrow \pi^- + \Xi_c^+$ &$4.4 \pm 1.6$ & $ 1.3 \pm 0.9 $ & $0.55 \pm 0.43$ \\
			\hline \hline
			$\Sigma_b^{*+} \rightarrow \pi^+ + \Lambda_b^0$ & $6.0 \pm 1.1$& $7.0 \pm 1.6 $ & $0.74 \pm 0.21$ \\
			\hline
			$\Sigma_b^{*-} \rightarrow \pi^- + \Lambda_b^0$ & $6.0 \pm 1.1$& $7.8 \pm 2.9$ & $0.75 \pm 0.34$\\
			\toprule
			\midrule
			\hline \hline
			$\Xi_b^{*0} \rightarrow \pi^0 + \Xi_b^0$ & $7.5 \pm 2.6$& $1.4 \pm 0.9$ &$ 1.0 \pm 0.7 $  \\
			\hline
			$\Xi_b^{*0} \rightarrow \pi^+ + \Xi_b^-$ &$10.6 \pm 3.7$ & $ 1.9 \pm 1.3 $ &$ 1.0 \pm 0.5 $ \\
			\hline
			$\Xi_b^{*-} \rightarrow \pi^0 + \Xi_b^-$ & $ 7.5\pm 2.6 $& $1.7 \pm 1.2$ & $ 1.0 \pm 0.6 $ \\
			\hline
			$\Xi_b^{*-} \rightarrow \pi^- + \Xi_b^0$ &$10.6 \pm 3.7$ & $ 3.7 \pm 2.5 $ &  $ 1.0 \pm 0.7 $\\
		\end{tabular}
	\end{minipage}
	\caption{The strong couplings, and the corresponding decay widths and branching fractions for the allowed strong decays of ${\cal B}_1 \rightarrow {\cal B}_2 {\cal M}$ for $\Delta S=1$.}
	\label{tab:decay_channels}
\end{table*}

\textit{\textbf{\textcolor{violet}{Conclusions:}}~}
We considered  $ 115000 $ potentially possible strong decay scenarios  of well-established baryons and mesons: Decays of  an initial baryon  with spin 3/2 or 1/2  to  a final baryon with spin 3/2 or 1/2 $ + $ a meson (pseudoscalar and vector) existing in PDG.  Then we applied  some restrictions and conservation laws:   The kinematical energy-momentum conservation,  conservation of the  total angular momentum,  and the  condition pertains to the quark structure of the resulting hadrons based on the fact that strong decays do not change the flavor of the quarks involved. The last condition means that  if the quark content of the initial and final baryons is the same, 
the meson should have the structure $q\bar{q}$ (the same quark-antiquark).  However, 
if the two-baryon structures are different, the baryons must have two identical quarks, 
and the quark structure of the meson inherits different quarks: a quark from the  initial baryon and an antiquark with the same flavor as the final baryon.  The remaining channels after applying the above constrains were categorized into $\Delta S=0$ and $\Delta S=1$ configurations. By utilizing a \texttt{Mathematica}/\texttt{Python} code, the kinematically allowed strong decays were identified, resulting in five decays in the $\Delta S=0$ category and 33 decays in the $\Delta S=1$ category. The corresponding decay widths and branching fractions were subsequently calculated.
 The results are found to be consistent with  expectations and the existing data provided by PDG for some channels. For those channels that the experimental results are not separated considering the charges and flavors of the participating particles in PDG,  when the contributions from all relevant isospin components in our case are properly summed,  our results demonstrate a meaningful and qualitatively consistent agreement with the PDG values, within the stated uncertainties. Future high-resolution experimental analyses in charge-specific channels could provide a more rigorous test and validation of our detailed predictions.
   The obtained results can help the ongoing theoretical and experimental research on the strong decays of baryons.

\section*{ACKNOWLEDGMENTS}

 K.  Azizi thanks  Iran national science foundation (INSF)
for the partial financial support provided under the elites Grant No. 4037888.  K.  Azizi and S.  Rostami are  grateful to the CERN-TH division for their support and warm hospitality.

	\onecolumngrid

	\twocolumngrid

	\onecolumngrid


\begin{thebibliography}{99}
		
\bibitem{Yuan:2021eqz}
C.~Z.~Yuan,
``New experimental results on light and heavy hadrons,''
\href{https://doi.org/10.22323/1.380.0016}{PoS \textbf{PANIC2021}, 016 (2022)},
\href{https://arxiv.org/abs/2111.07514}{[arXiv:2111.07514 [hep-ex]].}


\bibitem{Rossi:2023xqa}
A.~Rossi,
``Hadronization mechanism (via heavy-flavor hadrons): Experiment,''
\href{https://doi.org/10.22323/1.438.0022}{PoS \textbf{HardProbes2023}, 022 (2024)},
\href{https://arxiv.org/abs/2308.10202}{[arXiv:2308.10202 [hep-ex]].}

\bibitem{Barnes:2003bn}
T.~Barnes,
``Strong decays: Past, present and future,''
\href{https://doi.org/10.1063/1.1799772}{AIP Conf. Proc. \textbf{717}, no.1, 625-635 (2004)},
\href{https://arxiv.org/abs/0311102}{[arXiv:hep-ph/0311102 [hep-ph]].}


\bibitem{Akita:2024nam}
K.~Akita, G.~Baur, M.~Ovchynnikov, T.~Schwetz and V.~Syvolap,
``New physics decaying into metastable particles: impact on cosmic neutrinos,''
\href{https://arxiv.org/abs/2411.00892}{[arXiv:2411.00892 [hep-ph]].}

\bibitem{Alford:1998mk}
M.~G.~Alford, K.~Rajagopal and F.~Wilczek,
``Color flavor locking and chiral symmetry breaking in high density QCD,''
\href{https://doi.org/10.1016/S0550-3213(98)00668-3}{Nucl. Phys. B \textbf{537}, 443-458 (1999),}
\href{https://arxiv.org/abs/hep-ph/9804403}{[arXiv:hep-ph/9804403 [hep-ph]].}

\bibitem{Matsui:1986dk}
T.~Matsui and H.~Satz,
``$J/\psi$ Suppression by Quark-Gluon Plasma Formation,''
\href{https://doi.org/10.1016/0370-2693(86)91404-8}{Phys. Lett. B \textbf{178}, 416-422 (1986).}

\bibitem{Kohri:2001jx}
K.~Kohri,
``Primordial nucleosynthesis and hadronic decay of a massive particle with a relatively short lifetime,''
\href{https://doi.org/10.1103/PhysRevD.64.043515}{Phys. Rev. D \textbf{64}, 043515 (2001),}
\href{https://arxiv.org/abs/astro-ph/0103411}{[arXiv:astro-ph/0103411 [astro-ph]].}



\bibitem{Tawfiq:1998nk}
S.~Tawfiq, P.~J.~O'Donnell and J.~G.~Korner,
``Charmed baryon strong coupling constants in a light front quark model,''
\href{https://doi.org/10.1103/PhysRevD.58.054010}{Phys. Rev. D \textbf{58}, 054010 (1998)},
\href{https://arxiv.org/abs/9803246}{[arXiv:hep-ph/9803246 [hep-ph]].}


\bibitem{Ivanov:1999bk}
M.~A.~Ivanov, J.~G.~Korner, V.~E.~Lyubovitskij and A.~G.~Rusetsky,
``Strong and radiative decays of heavy flavored baryons,''
\href{https://doi.org/10.1103/PhysRevD.60.094002}{Phys. Rev. D \textbf{60}, 094002 (1999)},
\href{https://arxiv.org/abs/9904421}{[arXiv:hep-ph/9904421 [hep-ph]].}


\bibitem{Azizi:2008ui}
K.~Azizi, M.~Bayar and A.~Ozpineci,
``Sigma(Q) Lambda(Q) pi Coupling Constant in Light Cone QCD Sum Rules,''
\href{https://doi.org/10.1103/PhysRevD.79.056002}{Phys. Rev. D \textbf{79}, 056002 (2009)},
\href{https://arxiv.org/abs/0811.2695}{[arXiv:0811.2695 [hep-ph]].}


\bibitem{Aliev:2009ei}
T.~M.~Aliev, A.~Ozpineci, M.~Savci and V.~S.~Zamiralov,
``Vector meson-baryon strong coupling contants in light cone QCD sum rules,''
\href{https://doi.org/10.1103/PhysRevD.80.016010}{Phys. Rev. D \textbf{80}, 016010 (2009)},
\href{https://arxiv.org/abs/0905.4664}{[arXiv:0905.4664 [hep-ph]].}


\bibitem{Huang:2009is}
P.~Z.~Huang, H.~X.~Chen and S.~L.~Zhu,
``Light vector meson and heavy baryon strong interaction,''
\href{https://doi.org/10.1103/PhysRevD.80.094007}{Phys. Rev. D \textbf{80}, 094007 (2009)},
\href{https://arxiv.org/abs/0909.5551}{[arXiv:0909.5551 [hep-ph]].}



\bibitem{Aliev:2010yx}
T.~M.~Aliev, K.~Azizi and M.~Savci,
``Strong coupling constants of light pseudoscalar mesons with heavy baryons in QCD,''
\href{https://doi.org/10.1016/j.physletb.2010.12.027}{Phys. Lett. B \textbf{696}, 220-226 (2011)},
\href{https://arxiv.org/abs/1009.3658}{[arXiv:1009.3658 [hep-ph]].}

\bibitem{Aliev:2010ev}
T.~M.~Aliev, K.~Azizi and M.~Savci,
``Spin--3/2 to spin--1/2 heavy baryons and pseudoscalar mesons transitions in QCD,''
\href{https://doi.org/10.1140/epjc/s10052-011-1675-5}{Eur. Phys. J. C \textbf{71}, 1675 (2011)},
\href{https://arxiv.org/abs/1012.5935}{[arXiv:1012.5935 [hep-ph]].}


\bibitem{Aliev:2010dw}
T.~M.~Aliev, K.~Azizi and M.~Savci,
``Strong Coupling Constants of Decuplet Baryons with Vector Mesons,''
\href{https://doi.org/10.1103/PhysRevD.82.096006}{Phys. Rev. D \textbf{82}, 096006 (2010)},
\href{https://arxiv.org/abs/1007.3389}{[arXiv:1007.3389 [hep-ph]].}

\bibitem{Aliev:2010su}
T.~M.~Aliev, K.~Azizi and M.~Savci,
``Strong transitions of decuplet to octet baryons and pseudoscalar mesons,''
\href{https://doi.org/10.1016/j.nuclphysa.2010.06.013}{Nucl. Phys. A \textbf{847}, 101-117 (2010)},
\href{https://arxiv.org/abs/1003.5467 }{[arXiv:1003.5467 [hep-ph]].}


\bibitem{Azizi:2010sy}
K.~Azizi, M.~Bayar, A.~Ozpineci and Y.~Sarac,
``The $g_{\Sigma_Q\Sigma_Q\pi}$ Coupling Constant via Light Cone QCD Sum Rules,''
\href{https://doi.org/10.1103/PhysRevD.82.076004}{Phys. Rev. D \textbf{82}, 076004 (2010)},
\href{https://arxiv.org/abs/1008.0202 }{[arXiv:1008.0202 [hep-ph]].}



\bibitem{Detmold:2011bp}
W.~Detmold, C.~J.~D.~Lin and S.~Meinel,
``Axial couplings and strong decay widths of heavy hadrons,''
\href{https://doi.org/10.1103/PhysRevLett.108.172003}{Phys. Rev. Lett. \textbf{108}, 172003 (2012)},
\href{https://arxiv.org/abs/1109.2480 }{[arXiv:1109.2480 [hep-lat]].}


\bibitem{Aliev:2011ufa}
T.~M.~Aliev, K.~Azizi and M.~Savci,
``Strong coupling constants of heavy spin--3/2 baryons with light pseudoscalar mesons,''
\href{https://doi.org/10.1016/j.nuclphysa.2011.08.007}{Nucl. Phys. A \textbf{870-871}, 58-71 (2011)},
\href{https://arxiv.org/abs/1102.5460}{[arXiv:1102.5460 [hep-ph]].}


\bibitem{Azizi:2014bua}
K.~Azizi, Y.~Sarac and H.~Sundu,
``Strong $\Lambda_bNB$ and $\Lambda_cND$ vertices,''
\href{https://doi.org/10.1103/PhysRevD.90.114011}{Phys. Rev. D \textbf{90}, no.11, 114011 (2014)},
\href{https://arxiv.org/abs/1410.7548}{[arXiv:1410.7548 [hep-ph]].}


\bibitem{Azizi:2015bxa}
K.~Azizi, Y.~Sarac and H.~Sundu,
``Strong couplings of negative and positive parity nucleons to the heavy baryons and mesons,''
\href{https://doi.org/10.1103/PhysRevD.92.014022}{Phys. Rev. D \textbf{92}, no.1, 014022 (2015)},
\href{https://arxiv.org/abs/1506.00809}{[arXiv:1506.00809 [hep-ph]].}


\bibitem{Azizi:2015jya}
K.~Azizi, Y.~Sarac and H.~Sundu,
``On the strong coupling N$^{(*)}$N$^{(*)}$ $\pi$,''
\href{https://doi.org/10.1140/epja/i2016-16114-2}{Eur. Phys. J. A \textbf{52}, no.4, 114 (2016)},
\href{https://arxiv.org/abs/1510.05432}{[arXiv:1510.05432 [hep-ph]].}


\bibitem{Azizi:2015tya}
K.~Azizi, Y.~Sarac and H.~Sundu,
``Strong $\Sigma_bNB$ and $\Sigma_cND$ coupling constants in QCD,''
\href{https://doi.org/10.1016/j.nuclphysa.2015.09.005}{Nucl. Phys. A \textbf{943}, 159-167 (2015)},
\href{https://arxiv.org/abs/1501.05084}{[arXiv:1501.05084 [hep-ph]].}


\bibitem{Agaev:2015faa}
S.~S.~Agaev, K.~Azizi and H.~Sundu,
``Strong $D^{*}_sD_{s}\eta^{(\prime)}$ and $B^{*}_sB_{s}\eta^{(\prime)}$ vertices from QCD light-cone sum rules,''
\href{https://doi.org/10.1103/PhysRevD.92.116010}{Phys. Rev. D \textbf{92}, no.11, 116010 (2015)},
\href{https://arxiv.org/abs/1509.08620}{[arXiv:1509.08620 [hep-ph]].}


\bibitem{Aliev:2016qdo}
T.~M.~Aliev and M.~Savc\i{},
``Strong coupling constant of negative parity octet baryons with light pseudoscalar mesons in light cone QCD sum rules,''
\href{https://doi.org/10.1016/j.nuclphysa.2016.10.005}{Nucl. Phys. A \textbf{957}, 391-405 (2017)},
\href{https://arxiv.org/abs/1610.03979}{[arXiv:1610.03979 [hep-ph]].}


\bibitem{Aliev:2017tej}
T.~M.~Aliev, S.~Bilmis and M.~Savci,
``Strong Coupling Constants of Negative Parity Heavy Baryons with $\pi$ and $K$ Mesons,''
\href{https://doi.org/10.1155/2017/2493140}{Adv. High Energy Phys. \textbf{2017}, 2493140 (2017)},
\href{https://arxiv.org/abs/1712.01574}{[arXiv:1712.01574 [hep-ph]].}


\bibitem{Yu:2018hnv}
G.~L.~Yu, R.~H.~Guan and Z.~G.~Wang,
``Analysis of the strong vertices of $\Sigma_cND^{*}$ and $\Sigma_bNB^{*}$ in QCD sum rules,''
\href{https://doi.org/10.1142/S0217751X18502172}{Int. J. Mod. Phys. A \textbf{33}, no.36, 1850217 (2019)},
\href{https://arxiv.org/abs/1810.05970 }{[arXiv:1810.05970 [hep-ph]].}


\bibitem{Alrebdi:2020rev}
H.~I.~Alrebdi, T.~M.~Aliev and K.~\c{S}im\c{s}ek,
``Determination of the strong vertices of doubly heavy baryons with pseudoscalar mesons in QCD,''
\href{https://doi.org/10.1103/PhysRevD.102.074007}{Phys. Rev. D \textbf{102}, no.7, 074007 (2020)},
\href{https://arxiv.org/abs/2008.05098}{[arXiv:2008.05098 [hep-ph]].}


\bibitem{Rostami:2020euc}
S.~Rostami, K.~Azizi and A.~R.~Olamaei,
``Strong Coupling Constants of the Doubly Heavy Spin-1/2 Baryons with Light Pseudoscalar Mesons,''
\href{https://doi.org/10.1088/1674-1137/abd084}{Chin. Phys. C \textbf{45}, no.2, 023120 (2021)},
\href{https://arxiv.org/abs/2008.12715 }{[arXiv:2008.12715 [hep-ph]].}


\bibitem{Aliev:2020aon}
T.~M.~Aliev and K.~\c{S}im\c{s}ek,
``Strong coupling constants of doubly heavy baryons with vector mesons in QCD,''
\href{https://doi.org/10.1140/epjc/s10052-020-08553-z}{Eur. Phys. J. C \textbf{80}, no.10, 976 (2020)},
\href{https://arxiv.org/abs/2009.03464}{[arXiv:2009.03464 [hep-ph]].}


\bibitem{Aliev:2020lly}
T.~M.~Aliev and K.~\c{S}im\c{s}ek,
``Strong vertices of doubly heavy spin- $3/2$ \textendash{}spin- $1/2$ baryons with light mesons in light-cone QCD sum rules,''
\href{https://doi.org/10.1103/PhysRevD.103.054044}{Phys. Rev. D \textbf{103}, no.5, 054044 (2021)},
\href{https://arxiv.org/abs/2011.07150}{[arXiv:2011.07150 [hep-ph]].}


\bibitem{Azizi:2020zin}
K.~Azizi, A.~R.~Olamaei and S.~Rostami,
``Strong interaction of doubly heavy spin-3/2 baryons with light vector mesons,''
\href{https://doi.org/10.1140/epjc/s10052-020-08770-6}{Eur. Phys. J. C \textbf{80}, no.12, 1196 (2020)},
\href{https://arxiv.org/abs/2011.02919}{[arXiv:2011.02919 [hep-ph]].}


\bibitem{Olamaei:2021hjd}
A.~R.~Olamaei, K.~Azizi and S.~Rostami,
``Strong vertices of doubly heavy spin-3/2 baryons with light pseudoscalar mesons,''
\href{https://doi.org/10.1088/1674-1137/ac224b}{Chin. Phys. C \textbf{45}, no.11, 113107 (2021)},
\href{https://arxiv.org/abs/2102.03852}{[arXiv:2102.03852 [hep-ph]].}


\bibitem{Aliev:2021hqq}
T.~M.~Aliev, T.~Barakat and K.~\c{S}im\c{s}ek,
``Strong $ B_{QQ'}^* B_{QQ'} V $ vertices and the radiative decays of $ B_{QQ}^* \to B_{QQ} \gamma $ in the light-cone sum rules,''
\href{https://doi.org/10.1140/epja/s10050-021-00471-2}{Eur. Phys. J. A \textbf{57}, no.5, 160 (2021)},
\href{https://arxiv.org/abs/2101.10264}{[arXiv:2101.10264 [hep-ph]].}


\bibitem{Lu:2023pcg}
J.~Lu, G.~L.~Yu, Z.~G.~Wang and B.~Wu,
``Analysis of the strong vertices of $\Sigma _{c}\Delta D^{*}$ and $\Sigma _{b}\Delta B^{*}$ in QCD sum rules,''
\href{https://doi.org/10.1140/epjc/s10052-023-12076-8}{Eur. Phys. J. C \textbf{83}, no.10, 907 (2023)},
\href{https://arxiv.org/abs/2308.06705}{[arXiv:2308.06705 [hep-ph]].}


\bibitem{Bahtiyar:2020uuj}
H.~Bahtiyar, K.~U.~Can, G.~Erkol, P.~Gubler, M.~Oka and T.~T.~Takahashi,
``Charmed baryon spectrum from lattice QCD near the physical point,''
\href{https://doi.org/10.1103/PhysRevD.102.054513}{Phys. Rev. D \textbf{102}, no.5, 054513 (2020),}
\href{https://arxiv.org/abs/2004.08999}{[arXiv:2004.08999 [hep-lat]].}

\bibitem{Zhang:2021oja}
Q.~A.~Zhang, J.~Hua, F.~Huang, R.~Li, Y.~Li, C.~L\"u, C.~D.~Lu, P.~Sun, W.~Sun and W.~Wang, \textit{et al.}
``First lattice QCD calculation of semileptonic decays of charmed-strange baryons $\Xi_{c}$*,''
\href{https://doi.org/10.1088/1674-1137/ac2b12}{Chin. Phys. C \textbf{46}, no.1, 011002 (2022),}
\href{https://arxiv.org/abs/2103.07064}{[arXiv:2103.07064 [hep-lat]].}

\bibitem{Agaev:2017ywp}
S.~S.~Agaev, K.~Azizi and H.~Sundu,
``Decay widths of the excited $\Omega_b$ baryons,''
\href{https://doi.org/10.1103/PhysRevD.96.094011}{Phys. Rev. D \textbf{96}, no.9, 094011 (2017),}
\href{https://arxiv.org/abs/1708.07348}{[arXiv:1708.07348 [hep-ph]].}

\bibitem{Aliev:2018vye}
T.~M.~Aliev, K.~Azizi, Y.~Sarac and H.~Sundu,
``Determination of the quantum numbers of $\Sigma_b(6097)^{\pm}$ via their strong decays,''
\href{https://doi.org/10.1103/PhysRevD.99.094003}{Phys. Rev. D \textbf{99}, no.9, 094003 (2019),}
\href{https://arxiv.org/abs/1811.05686}{[arXiv:1811.05686 [hep-ph]].}

\bibitem{Aliev:2018lcs}
T.~M.~Aliev, K.~Azizi, Y.~Sarac and H.~Sundu,
``Structure of the $\Xi_b(6227)^-$ resonance,''
\href{https://doi.org/10.1103/PhysRevD.98.094014}{Phys. Rev. D \textbf{98}, no.9, 094014 (2018),}
\href{https://arxiv.org/abs/1808.08032}{[arXiv:1808.08032 [hep-ph]].}

\bibitem{Yang:2022oog}
H.~M.~Yang, H.~X.~Chen, E.~L.~Cui and Q.~Mao,
``Identifying the $\Xi_b(6100)$ as the P-wave bottom baryon of $J^P=3/2^-$,''
\href{https://doi.org/10.1103/PhysRevD.106.036018}{Phys. Rev. D \textbf{106}, no.3, 036018 (2022),}
\href{https://arxiv.org/abs/2205.07224}{[arXiv:2205.07224 [hep-ph]].}

\bibitem{Agaev:2017lip}
S.~S.~Agaev, K.~Azizi and H.~Sundu,
``Interpretation of the new $\Omega_c^{0}$ states via their mass and width,''
\href{https://doi.org/10.1140/epjc/s10052-017-4953-z}{Eur. Phys. J. C \textbf{77}, no.6, 395 (2017),}
\href{https://arxiv.org/abs/1704.04928}{[arXiv:1704.04928 [hep-ph]].}

\bibitem{Vishwakarma:2022vzy}
K.~K.~Vishwakarma and A.~Upadhyay,
``Analysis of 2S singly heavy baryons in HQET,''
\href{https://doi.org/10.1142/S0217751X24500970}{Int. J. Mod. Phys. A \textbf{39}, no.25, 2450097 (2024),}
\href{https://arxiv.org/abs/2208.02536}{[arXiv:2208.02536 [hep-ph]].}

\bibitem{Jia:2019bkr}
D.~Jia, W.~N.~Liu and A.~Hosaka,
``Regge behaviors in orbitally excited spectroscopy of charmed and bottom baryons,''
\href{https://doi.org/10.1103/PhysRevD.101.034016}{Phys. Rev. D \textbf{101}, no.3, 034016 (2020),}
\href{https://arxiv.org/abs/1907.04958}{[arXiv:1907.04958 [hep-ph]].}

\bibitem{Oudichhya:2023awb}
J.~Oudichhya and A.~K.~Rai,
``Spin--parity identification of newly observed singly charmed baryons in Regge phenomenology,''
\href{https://doi.org/10.1140/epja/s10050-023-01024-5}{Eur. Phys. J. A \textbf{59}, no.6, 123 (2023).}

\bibitem{Karliner:2017kfm}
M.~Karliner and J.~L.~Rosner,
``Very narrow excited $\Omega_c$ baryons,''
\href{https://doi.org/10.1103/PhysRevD.95.114012}{Phys. Rev. D \textbf{95}, no.11, 114012 (2017),}
\href{https://arxiv.org/abs/1703.07774}{[arXiv:1703.07774 [hep-ph]].}

\bibitem{Wang:2017kfr}
K.~L.~Wang, Y.~X.~Yao, X.~H.~Zhong and Q.~Zhao,
``Strong and radiative decays of the low-lying $S$- and $P$-wave singly heavy baryons,''
\href{https://doi.org/10.1103/PhysRevD.96.116016}{Phys. Rev. D \textbf{96}, no.11, 116016 (2017),}
\href{https://arxiv.org/abs/1709.04268}{[arXiv:1709.04268 [hep-ph]].}

\bibitem{Yang:2017qan}
G.~Yang, J.~Ping and J.~Segovia,
``The $S$- and $P$-Wave Low-Lying Baryons in the Chiral Quark Model,''
\href{https://doi.org/10.1007/s00601-018-1433-4}{Few Body Syst. \textbf{59}, no.6, 113 (2018),}
\href{https://arxiv.org/abs/1709.09315}{[arXiv:1709.09315 [hep-ph]].}

\bibitem{Karliner:2018bms}
M.~Karliner and J.~L.~Rosner,
``Scaling of $P$-wave excitation energies in heavy-quark systems,''
\href{https://doi.org/10.1103/PhysRevD.98.074026}{Phys. Rev. D \textbf{98}, no.7, 074026 (2018),}
\href{https://arxiv.org/abs/1808.07869}{[arXiv:1808.07869 [hep-ph]].}

\bibitem{Shi:2019tji}
S.~Shi, J.~Zhao and P.~Zhuang,
``Heavy flavor dissociation in framework of multi-body Dirac equations,''
\href{https://doi.org/10.1088/1674-1137/44/8/084101}{Chin. Phys. C \textbf{44}, no.8, 084101 (2020),}
\href{https://arxiv.org/abs/1905.10627}{[arXiv:1905.10627 [nucl-th]].}

\bibitem{Karliner:2020fqe}
M.~Karliner and J.~L.~Rosner,
``Interpretation of excited $\Omega_b$ signals,''
\href{https://doi.org/10.1103/PhysRevD.102.014027}{Phys. Rev. D \textbf{102}, no.1, 014027 (2020),}
\href{https://arxiv.org/abs/2005.12424}{[arXiv:2005.12424 [hep-ph]].}

\bibitem{Wang:2020gkn}
K.~L.~Wang, L.~Y.~Xiao and X.~H.~Zhong,
``Understanding the newly observed $\Xi_c^0$ states through their decays,''
\href{https://doi.org/10.1103/PhysRevD.102.034029}{Phys. Rev. D \textbf{102}, no.3, 034029 (2020),}
\href{https://arxiv.org/abs/2004.03221}{[arXiv:2004.03221 [hep-ph]].}

\bibitem{Xiao:2020gjo}
L.~Y.~Xiao and X.~H.~Zhong,
``Toward establishing the low-lying $P$-wave $\Sigma_b$ states,''
\href{https://doi.org/10.1103/PhysRevD.102.014009}{Phys. Rev. D \textbf{102}, no.1, 014009 (2020),}
\href{https://arxiv.org/abs/2004.11106}{[arXiv:2004.11106 [hep-ph]].}

\bibitem{Chen:2021eyk}
B.~Chen, S.~Q.~Luo and X.~Liu,
``Universal behavior of mass gaps existing in the single heavy baryon family,''
\href{https://doi.org/10.1140/epjc/s10052-021-09234-1}{Eur. Phys. J. C \textbf{81}, no.5, 474 (2021),}
\href{https://arxiv.org/abs/2101.10806}{[arXiv:2101.10806 [hep-ph]].}

\bibitem{Garcia-Tecocoatzi:2022zrf}
H.~Garcia-Tecocoatzi, A.~Giachino, J.~Li, A.~Ramirez-Morales and E.~Santopinto,
``Strong decay widths and mass spectra of charmed baryons,''
\href{https://doi.org/10.1103/PhysRevD.107.034031}{Phys. Rev. D \textbf{107}, no.3, 034031 (2023),}
\href{https://arxiv.org/abs/2205.07049}{[arXiv:2205.07049 [hep-ph]].}

\bibitem{Ma:2022vqf}
Y.~Ma, L.~Meng, Y.~K.~Chen and S.~L.~Zhu,
``Ground state baryons in the flux-tube three-body confinement model using diffusion Monte~Carlo,''
\href{https://doi.org/10.1103/PhysRevD.107.054035}{Phys. Rev. D \textbf{107}, no.5, 054035 (2023),}
\href{https://arxiv.org/abs/2211.09021}{[arXiv:2211.09021 [hep-ph]].}

\bibitem{Wang:2022dmw}
W.~J.~Wang, L.~Y.~Xiao and X.~H.~Zhong,
``Strong decays of the low-lying $\rho$-mode $1P$-wave singly heavy baryons,''
\href{https://doi.org/10.1103/PhysRevD.106.074020}{Phys. Rev. D \textbf{106}, no.7, 074020 (2022),}
\href{https://arxiv.org/abs/2208.10116}{[arXiv:2208.10116 [hep-ph]].}

\bibitem{Karliner:2023okv}
M.~Karliner and J.~L.~Rosner,
``Excited $\Omega_c$ Baryons as $2S$ states,''
\href{https://doi.org/10.1103/PhysRevD.108.014006}{Phys. Rev. D \textbf{108}, no.1, 014006 (2023),}
\href{https://arxiv.org/abs/2304.00407}{[arXiv:2304.00407 [hep-ph]].}

\bibitem{Ortiz-Pacheco:2023bns}
E.~Ortiz-Pacheco and R.~Bijker,
``Heavy and baryons in the quark model,''
\href{https://doi.org/10.1088/1742-6596/2619/1/012011}{J. Phys. Conf. Ser. \textbf{2619}, no.1, 012011 (2023),}
\href{https://arxiv.org/abs/2309.12266}{[arXiv:2309.12266 [hep-ph]].}

\bibitem{LHCb:2017uwr}
R.~Aaij \textit{et al.} [LHCb],
``Observation of five new narrow $\Omega_c^0$ states decaying to $\Xi_c^+ K^-$,''
\href{https://doi.org/10.1103/PhysRevLett.118.182001}{Phys. Rev. Lett. \textbf{118}, no.18, 182001 (2017),}
\href{https://arxiv.org/abs/1703.04639}{[arXiv:1703.04639 [hep-ex]].}

\bibitem{Belle:2017ext}
J.~Yelton \textit{et al.} [Belle],
``Observation of Excited $\Omega_c$ Charmed Baryons in $e^+e^-$ Collisions,''
\href{https://doi.org/10.1103/PhysRevD.97.051102}{Phys. Rev. D \textbf{97}, no.5, 051102 (2018),}
\href{https://arxiv.org/abs/1711.07927}{[arXiv:1711.07927 [hep-ex]].}

\bibitem{LHCb:2018haf}
R.~Aaij \textit{et al.} [LHCb],
``Observation of Two Resonances in the $\Lambda_b^0 \pi^\pm$ Systems and Precise Measurement of $\Sigma_b^\pm$ and $\Sigma_b^{*\pm}$ properties,''
\href{https://doi.org/10.1103/PhysRevLett.122.012001}{Phys. Rev. Lett. \textbf{122}, no.1, 012001 (2019),}
\href{https://arxiv.org/abs/1809.07752}{[arXiv:1809.07752 [hep-ex]].}

\bibitem{LHCb:2020tqd}
R.~Aaij \textit{et al.} [LHCb],
``First observation of excited $\Omega_b^-$ states,''
\href{https://doi.org/10.1103/PhysRevLett.124.082002}{Phys. Rev. Lett. \textbf{124}, no.8, 082002 (2020),}
\href{https://arxiv.org/abs/2001.00851}{[arXiv:2001.00851 [hep-ex]].}

\bibitem{LHCb:2021ssn}
R.~Aaij \textit{et al.} [LHCb],
``Observation of Two New Excited $\Xi_b^0$ States Decaying to $\Lambda^0_b K^- \pi^+$,''
\href{https://doi.org/10.1103/PhysRevLett.128.162001}{Phys. Rev. Lett. \textbf{128}, no.16, 162001 (2022),}
\href{https://arxiv.org/abs/2110.04497}{[arXiv:2110.04497 [hep-ex]].}

\bibitem{LHCb:2023tma}
R.~Aaij \textit{et al.} [LHCb],
``Observation and branching fraction measurement of the decay $\Xi_b^-\to\Lambda_b^0\pi^-$,''
\href{https://doi.org/10.1103/PhysRevD.108.072002}{Phys. Rev. D \textbf{108}, no.7, 072002 (2023),}
\href{https://arxiv.org/abs/2307.09427}{[arXiv:2307.09427 [hep-ex]].}



\bibitem{Shifman:1978bx}
M.~A.~Shifman, A.~I.~Vainshtein and V.~I.~Zakharov,
``QCD and Resonance Physics. Theoretical Foundations,''
\href{https://doi.org/10.1016/0550-3213(79)90022-1}{Nucl. Phys. B \textbf{147}, 385-447 (1979)}.

\bibitem{Colangelo:2000dp}
P.~Colangelo and A.~Khodjamirian,
``QCD sum rules, a modern perspective,''
\href{https://arxiv.org/abs/hep-ph/0010175}{[arXiv:hep-ph/0010175 [hep-ph]].}		

\bibitem{ParticleDataGroup:2024cfk}
S.~Navas \textit{et al.} [Particle Data Group],
``Review of particle physics,''
\href{https://doi.org/10.1103/PhysRevD.110.030001}{Phys. Rev. D \textbf{110}, no.3, 030001 (2024)}.

\end{thebibliography}
\end{document}